\documentclass[review]{elsarticle}

\usepackage{lineno,hyperref}
\modulolinenumbers[5]

\journal{arXiv}

\begin{document}

\begin{frontmatter}

\title{Entanglement between two qubits one of which interacts with a  thermal field}

\author{Eugene Bashkirov, Michail Mastyugin\corref{mycorrespondingauthor}}
\cortext[mycorrespondingauthor]{Corresponding author}
\ead{bash@samsu.ru}

\address{Department of General and Theoretical Physics,  Samara State University, 1 Academician Pavlov str., Samara 443011, Russia}

\begin{abstract} In this paper, we have investigated the entanglement between two dipole coupled two-level atoms. The model, in which  only one atom  is trapped in an lossless cavity and interacts with single-mode thermal field, and the other one can be spatially moved freely outside the cavity has been carried out. We have considered the effect of the atomic coherence on the entanglement behavior.  We have shown that a thermal field  might cause high entanglement between the atoms both for coherent and incoherent initial atomic states only for small values of the cavity mean photon number. In the considered model the atoms would  get entangled even when both atoms are initially in the excited state. We have also derived  that the degree of entanglement is weakly dependent on the strength of dipole-dipole interaction for coherent initial states.
\end{abstract}

\begin{keyword}
  Entanglement \sep Cavity QED \sep Negativity \sep Thermal field \sep Atomic coherence
\end{keyword}

\end{frontmatter}


\section{Introduction}
Entanglement between separate quantum systems is
one of the key problem in quantum mechanics. It plays a central role
in quantum information, quantum computation and communication,  and quantum cryptography  \cite{Nielsen}. Several
methods of creating entanglement have been proposed
involving trapped and cooled ions or neutral atoms in cavities, superconducting circuits, spins in solids etc. \cite{Buluta}. In order to function optimally various applications require  maximally entangled states.  Because of decoherence, which is generally related
to noise, there is great difficulty in generating and keeping
the integrity of a pure entangled states. Although
the interaction between the environment and quantum
systems can lead to decoherence, it may also be associated
with the formation of non-classical effects such as
entanglement  \cite{Plenio}. Thus, understanding and investigating entanglement of mixed states becomes one of the actual  problem of quantum information. Recently,  Bose et al.  \cite{Bose} have shown that entanglement can always arise in the interaction of an arbitrary large system in any
 mixed state with a single qubit in a pure  state,  and illustrated this using the Jaynes-Cummings interaction of a
 two-level atom in a pure state with a field in a thermal state at an
 arbitrary high temperature.   Kim et al. \cite{Kim} have investigated the atom-atom entanglement in the system of
  two identical   two-level atoms with one-photon  transition induced by a single-mode
 thermal field. They showed that a chaotic field with minimal information can entangled atoms which were
 prepared initially in a separable
 state. Zhang directly generalizes Kim's study to the case when
the atoms are slightly detuned from the thermal field, and study
how the detuning would affect atom-atom entanglement \cite{Zhang}.
  Zhou et al. \cite{Zhou2} have considered the same problem for nonidentical atoms with different couplings. The entanglement between
 two identical  two-level atoms through  nonlinear two-photon interaction with one-mode thermal field has been
 studied by Zhou  et al. \cite{Zhou1}. They  showed that atom-atom entanglement induced by nonlinear interaction is larger than that
 induced by linear interaction. In \cite{Bash1} has discovered that two atoms can be entangled also through nonlinear  nondegenerate two-photon interaction with two-mode thermal field. The influence of dipole-dipole interaction on entanglement between two cubits induced by one-mode and two-mode thermal field has been investigated in \cite{Aguiar}-\cite{Bash3}.

 The problem of creating or controlling the atomic entanglement is greatly related to the atomic coherence of population between different levels. The authors of papers \cite{A}-\cite{B} have shown that the entanglement between two atoms induced by one-mode or two-mode thermal field can be manipulated by changing the initial parameters of the atoms, such as the superposition coefficients and the relative phases of the initial atomic coherent state and the mean photon number of the cavity field.
  They have also discovered that  entanglement may be greatly enhanced due to dipole-dipole interaction in the presence of the atomic coherence \cite{Hu3}.

 While, as has been mentioned in \cite{Guo} the practical applications in quantum information processing require engineering entangled atoms so this expects operable atoms which  can be moved to distance without losing of information. Recently many schemes have been proposed to realize the engineering entangled atoms \cite{Marr}-\cite{Lu}. Guo and coauthors  have
 proposed a simple scheme to realize an easily engineered two-atom entangled
state. The advantage of this scheme is only one atom is trapped in a cavity, and
the other one can be spatially moved freely outside the cavity. But the authors have investigated the atom-atom entanglement only for vacuum initial cavity field.
 In this paper we study the   entanglement dynamics for model supposed in \cite{Guo} for thermal cavity field taking into account the initial atomic coherence.

\section{The model}
We consider two identical two-level atoms and one-mode quantum electromagnetic cavity field. The first atom is trapped in a lossless microcavity and resonantly
interacts with the cavity field of the frequency $\omega$. The second atom   lies beside the first atom  out of the cavity. We assume that the distance between atoms can compare with a wavelength on working transition. In this case the dipole-dipole interaction should be included.
The Hamiltonian of this system can be written as
$$ H=(1/2)\hbar \omega (\sigma _{1}^{z} +\sigma _{2}^{z} )+\hbar \omega a^{+} a + \hbar g(\sigma _{1}^{+} a+a^{+} \sigma _{1}^{-} )+\hbar J(\sigma _{1}^{+} \sigma _{2}^{-} +\sigma_{1}^{-} \sigma _{2}^{+} ),\eqno{(1)}
$$
 where  $(1/2)\sigma _{i}^{z} $ is the inversion operator for the $i$th atom ($i=1,2$), $\sigma _{i}^{+} =|+\rangle _{ii} \langle -|, $ and $ \sigma _{i}^{-} =|-\rangle _{ii} \langle +|$ are the transition operators between the excited  $|+\rangle _{i} $ and the ground  $|-\rangle _{i} $ states in the  $i$th atom, $a^+$ and $a$ are the creation and the annihilation operators of photons of  the cavity mode,   $g$ is the coupling constant
between atom and the cavity field and $J$ is the coupling
constant of the dipole interaction between the atoms and $|+\rangle$ and $|-\rangle$ are the excited and the ground states of
a single two-level atom. The two-atom wave function
can be expressed as a combination of state vectors of the form $|\it v_1,\it v_2\rangle = |\it v_1|| \it v_2\rangle$, where $\it v_1, v_2 =+,-.$

We consider that the initial states of atoms are the coherent superposition of the two levels, that is,
$$ |\Psi_1(0)\rangle  = \cos \theta_1 |+\rangle_1 + e^{\imath \varphi_1} \sin\theta_1 |-\rangle_1,$$
$$ |\Psi_2(0)\rangle  = \cos \theta_2 |+\rangle_1 + e^{\imath \varphi_2} \sin\theta_2 |-\rangle_2. \eqno{(2)}$$
Here  $\theta_1$ and $\theta_2$ denote the amplitudes of the polarized atoms, and $\varphi_1$ and $\varphi_2$ are relative phases of two atoms, respectively. So the initial density matrix for  two-atom system can be written as
$$\rho_A(0) = |\Psi_1(0)\rangle |\Psi_2(0)\rangle \langle \Psi_1(0)| \langle \Psi_2(0)|. $$

The initial cavity mode state are assumed to be the
thermal one-mode state
$$\rho_F(0)= \sum\limits_{n}  p_n |n\rangle \langle n|. $$
The weight functions are
$$p_n= \frac{{\bar n}^{n}}{(1+{\bar n})^{n+1}},$$
where ${\bar n_i}$  is the mean photon number in the $i$th cavity
mode, $${\bar n_i} = (\exp[\hbar \omega_i/k_BT]-1]^{-1},$$ $k_B$ is
the Boltzmann constant and $T$ is the equilibrium cavity temperature.

The initial density matrix of the whole system is
$$\rho(0) =\rho_A(0)  \rho_F(0) = \sum_n p_n |\Psi_1(0)\rangle |\Psi_2(0)\rangle \langle \Psi_1(0)| \langle \Psi_2(0)| |n\rangle \langle n|, \eqno{(3)}
$$

Before considering the dynamics of the  system  for thermal initial cavity field, it is straightforward to first study the case when the trapped two-level
atom  interacts with Fock state. Suppose that the excitation number of the atom-field system is $n$ ($n \geq 0$)
 the evolution of the system is confined in the subspace
$|-, -, n + 2\rangle, |+, -, n + 1\rangle, |-, +, n + 1\rangle,|+, +, n\rangle.$ On this basis, the  eigenfunctions  of the Hamiltonian  (1) can be written as \cite{Guo}
$$
|\Phi_{in}\rangle = \xi_{in} (X_{i1n} |-,-,n+2\rangle + X_{i2n} |+,-,n+1\rangle + $$
$$+ X_{i3n} |-,+,n+1\rangle + X_{i4n} |+,+,n\rangle)\quad (i=1,2,3,4),
$$
where
$$\xi_{in}= 1/\sqrt{|X_{i1n}|^2 +|X_{i2n}|^2 + |X_{i3n}|^2 + |X_{i4n}|^2}
$$
and
$$
X_{i1n} =1,\quad X_{i2n} =(-1)^{i+1} \frac{S_n}{\sqrt{n+2}}, $$
$$X_{i3n} = \frac{S_n^2-(n+2)}{(\sqrt{n+2})\alpha}, \quad X_{i4n} =(-1)^{i+1} \frac{S_n(S_n^2 -(n+2)-\alpha^2)}{(\sqrt{n+1}\sqrt{n+2}) \alpha}.
$$
Here $S_n = A_n$ for odd  $i$ and $S_n = B_n$ for even  $i$.
The corresponding eigenvalue are
$$
E_{1n}/\hbar= (n+1)\omega  + A_n, \qquad E_{2n}/\hbar=(n+1)\omega  - A_n,$$ $$  E_{3n}/\hbar=(n+1)\omega  + B_n, \qquad E_{4n}/\hbar=(n+1)\omega  - B_n.
$$
Here
$$
A_n=\sqrt{W_n+V_n}/2, \quad B_n=\sqrt{W_n-V_n}/2 $$ and $$ W_n= 4n+6+2\alpha^2,\quad V_n=2\sqrt{4(n+1)\alpha^2+(\alpha^2+1)},
$$
where $\alpha = J /g$.

To derive the full dynamics of our model one can consider also the basis states $|-,-,1\rangle, |+,-,0\rangle, |-,+,0\rangle$. In this basis the eigenfunctions and eigenvalues of the Hamiltonian (1)  are
$$
| \varphi_{1}\rangle = (\alpha^2/\Omega)[ |-,-,1\rangle - (1/\alpha) |-,+,0\rangle ], \qquad E_1 = 0; $$
$$| \varphi_{2}\rangle = (1/\sqrt{2})[ (1/\Omega)|-,-,1\rangle + |+,-,0\rangle +  (\alpha/\Omega) |-,+,0\rangle ], \qquad E_2/\hbar = \Omega; $$
$$| \varphi_{3}\rangle = (1/\sqrt{2})[- (1/\Omega)|-,-,1\rangle + |+,-,0\rangle -  (\alpha/\Omega) |-,+,0\rangle ], \qquad E_2/\hbar = - \Omega, $$
where $\Omega = \sqrt{1+\alpha^2}$.

At last, the Hamiltonian (1) has one more eigenfunction  $$\varphi_0 =|-,-,0\rangle $$
 which corresponds to energy $E_0 = -\hbar \omega$.

Assume that the whole system is initially in the state $|+,+,n\rangle$ $(n \geq 0$), then at time t, the whole system will evolve to
$$
|\Psi(t)\rangle = Z_{11,n} |-,-,n+2\rangle + Z_{21,n} |+,-,n+1\rangle  + Z_{31,n} |-,+,n+1\rangle  + Z_{41,n} |+,+,n\rangle. \eqno{(4)}$$
Here
$$
Z_{11,n} = e^{- \imath E_{1n} t/\hbar}\> \xi_{1n} \> Y_{41n} \> X_{11n} + e^{- \imath E_{2n} t/\hbar}\>  \xi_{2n}\> Y_{42n} X_{21n}+
$$
$$ + e^{- \imath E_{3n} t/\hbar}\> \xi_{3n}\> Y_{43n}\> X_{31n} + e^{- \imath E_{4n} t/\hbar} \>\xi_{4n}\> Y_{44n}\> X_{41n},$$
$$
Z_{21,n} = e^{- \imath E_{1n} t/\hbar}\> \xi_{1n}\> Y_{41n}\> X_{12n} + e^{- \imath E_{2n} t/\hbar]} \>\xi_{2n} \> Y_{42n}\> X_{22n}  + $$
$$+ e^{- \imath E_{3n} t/\hbar}\> \xi_{3n}\> Y_{43n}\> X_{32n} + e^{- \imath E_{4n} t/\hbar}\>  \xi_{4n} \> Y_{44n}\>  X_{42n},$$
$$
Z_{31,n} = e^{- \imath E_{1n} t/\hbar}\> \xi_{1n}\> Y_{41n}\> X_{13n} + e^{- \imath E_{2n} t/\hbar} \>\xi_{2n} \> Y_{42n}\> X_{23n}+ $$
$$ + e^{- \imath E_{3n} t/\hbar} \> \xi_{3n}\> Y_{43n}\> X_{33n} + e^{- \imath E_{4n} t/\hbar}\> \xi_{4n} \> Y_{44n}\> X_{43n},
$$
$$
Z_{41,n} = e^{-\imath E_{1n} t/\hbar}\> \xi_{1n}\> Y_{41n}\> X_{14n} + e^{- \imath E_{2n} t/\hbar}\> \xi_{2n}\> Y_{42n}\> X_{24n}+ $$
$$+ e^{- \imath E_{3n} t/\hbar}\> \xi_{3n}\> Y_{43n}\> X_{34n} + e^{- \imath E_{4n} t /\hbar}\> \xi_{4n}\> Y_{44n}\> X_{44n},$$
where $Y_{ijn}= \xi_{jn} X_{jin}^*$.

Similarly, when the whole system is initially in the state $|+,-,n+1\rangle$ $(n \geq 0$), then at time t, the whole system will evolve to
$$
|\Psi(t)\rangle = Z_{12,n} |-,-,n+2\rangle + Z_{22,n} |+,-,n+1\rangle  + Z_{32,n} |-,+,n+1\rangle  + Z_{42,n} |+,+,n\rangle. \eqno{(5)}$$
Here
$$
Z_{12,n} = e^{- \imath E_{1n} t/\hbar}\> \xi_{1n} \> Y_{21n} \> X_{11n} + e^{- \imath E_{2n} t/\hbar}\>  \xi_{2n}\> Y_{22n} X_{21n}+
$$
$$ + e^{- \imath E_{3n} t/\hbar}\> \xi_{3n}\> Y_{23n}\> X_{31n} + e^{- \imath E_{4n} t/\hbar} \>\xi_{4n}\> Y_{24n}\> X_{41n},$$
$$
Z_{22,n} = e^{- \imath E_{1n} t/\hbar}\> \xi_{1n}\> Y_{21n}\> X_{12n} + e^{- \imath E_{2n} t/\hbar]} \>\xi_{2n} \> Y_{22n}\> X_{22n}  + $$
$$+ e^{- \imath E_{3n} t/\hbar}\> \xi_{3n}\> Y_{23n}\> X_{32n} + e^{- \imath E_{4n} t/\hbar}\>  \xi_{4n} \> Y_{24n}\>  X_{42n},$$
$$
Z_{32,n} = e^{- \imath E_{1n} t/\hbar}\> \xi_{1n}\> Y_{21n}\> X_{13n} + e^{- \imath E_{2n} t/\hbar} \>\xi_{2n} \> Y_{22n}\> X_{23n}+ $$
$$ + e^{- \imath E_{3n} t/\hbar} \> \xi_{3n}\> Y_{23n}\> X_{33n} + e^{- \imath E_{4n} t/\hbar}\> \xi_{4n} \> Y_{24n}\> X_{43n},
$$
$$
Z_{42,n} = e^{-\imath E_{1n} t/\hbar}\> \xi_{1n}\> Y_{21n}\> X_{14n} + e^{- \imath E_{2n} t/\hbar}\> \xi_{2n}\> Y_{22n}\> X_{24n}+ $$
$$+ e^{- \imath E_{3n} t/\hbar}\> \xi_{3n}\> Y_{23n}\> X_{34n} + e^{- \imath E_{4n} t /\hbar}\> \xi_{4n}\> Y_{24n}\> X_{44n}.$$

If the initial state of considered system is $|+,-,0\rangle$, the time dependent wave function  takes the form
$$|\Psi(t)\rangle = Z_{12} |-,-,1\rangle + Z_{22} |+,-,0\rangle  + Z_{32} |-,+,0\rangle, \eqno{(6)}
$$
where
$$
Z_{12} = - \frac{\imath}{ \Omega} \sin\Omega t, \quad Z_{22} = \cos\Omega t, \quad Z_{32} = - \frac{\imath \alpha}{ \Omega} \sin\Omega t.
$$

For initial state $|-,+,n+1\rangle \> (n \geq 0$) the time-dependent wave function is
$$
|\Psi(t)\rangle = Z_{13,n} |-,-,n+2\rangle + Z_{23,n} |+,-,n+1\rangle  + Z_{33,n} |-,+,n+1\rangle  + Z_{43,n} |+,+,n\rangle. \eqno{(7)}$$
Here
$$
Z_{13,n} = e^{- \imath E_{1n} t/\hbar}\> \xi_{1n} \> Y_{31n} \> X_{11n} + e^{- \imath E_{2n} t/\hbar}\>  \xi_{2n}\> Y_{32n} X_{21n}+
$$
$$ + e^{- \imath E_{3n} t/\hbar}\> \xi_{3n}\> Y_{33n}\> X_{31n} + e^{- \imath E_{4n} t/\hbar} \>\xi_{4n}\> Y_{34n}\> X_{41n},$$
$$
Z_{23,n} = e^{- \imath E_{1n} t/\hbar}\> \xi_{1n}\> Y_{31n}\> X_{12n} + e^{- \imath E_{2n} t/\hbar]} \>\xi_{2n} \> Y_{32n}\> X_{22n}  + $$
$$+ e^{- \imath E_{3n} t/\hbar}\> \xi_{3n}\> Y_{33n}\> X_{32n} + e^{- \imath E_{4n} t/\hbar}\>  \xi_{4n} \> Y_{34n}\>  X_{42n},$$
$$
Z_{32,n} = e^{- \imath E_{1n} t/\hbar}\> \xi_{1n}\> Y_{31n}\> X_{13n} + e^{- \imath E_{2n} t/\hbar} \>\xi_{2n} \> Y_{32n}\> X_{23n}+ $$
$$ + e^{- \imath E_{3n} t/\hbar} \> \xi_{3n}\> Y_{33n}\> X_{33n} + e^{- \imath E_{4n} t/\hbar}\> \xi_{4n} \> Y_{34n}\> X_{43n},
$$
$$
Z_{42,n} = e^{-\imath E_{1n} t/\hbar}\> \xi_{1n}\> Y_{31n}\> X_{14n} + e^{- \imath E_{2n} t/\hbar}\> \xi_{2n}\> Y_{32n}\> X_{24n}+ $$
$$+ e^{- \imath E_{3n} t/\hbar}\> \xi_{3n}\> Y_{33n}\> X_{34n} + e^{- \imath E_{4n} t /\hbar}\> \xi_{4n}\> Y_{34n}\> X_{44n}.$$

If the initial state of considered system is $|-,+,0\rangle$, the time dependent wave function  of the whole system takes the form
$$|\Psi(t)\rangle = Z_{13} |-,-,1\rangle + Z_{23} |+,-,0\rangle  + Z_{33} |-,+,0\rangle, \eqno{(8)}
$$
where
$$
Z_{13} = - \frac{\alpha}{ \Omega^2} (\cos\Omega t -1), \quad Z_{23} = -\frac{\imath \alpha}{\Omega}\sin\Omega t, \quad Z_{33} = - \frac{1}{ \Omega^2} (1+\alpha^2\cos\Omega t).
$$

For initial  state $|-,-,n+2\rangle$ $(n \geq 0$) we have at time t that
$$
|\Psi(t)\rangle = Z_{14,n} |-,-,n+2\rangle + Z_{24,n} |+,-,n+1\rangle  + Z_{34,n} |-,+,n+1\rangle  + Z_{44,n} |+,+,n\rangle. \eqno{(9)}$$
Here
$$
Z_{14,n} = e^{- \imath E_{1n} t/\hbar}\> \xi_{1n} \> Y_{41n} \> X_{11n} + e^{- \imath E_{2n} t/\hbar}\>  \xi_{2n}\> Y_{42n} X_{21n}+
$$
$$ + e^{- \imath E_{3n} t/\hbar}\> \xi_{3n}\> Y_{43n}\> X_{31n} + e^{- \imath E_{4n} t/\hbar} \>\xi_{4n}\> Y_{44n}\> X_{41n},$$
$$
Z_{24,n} = e^{- \imath E_{1n} t/\hbar}\> \xi_{1n}\> Y_{41n}\> X_{12n} + e^{- \imath E_{2n} t/\hbar]} \>\xi_{2n} \> Y_{42n}\> X_{22n}  + $$
$$+ e^{- \imath E_{3n} t/\hbar}\> \xi_{3n}\> Y_{43n}\> X_{32n} + e^{- \imath E_{4n} t/\hbar}\>  \xi_{4n} \> Y_{44n}\>  X_{42n},$$
$$
Z_{42,n} = e^{- \imath E_{1n} t/\hbar}\> \xi_{1n}\> Y_{41n}\> X_{13n} + e^{- \imath E_{2n} t/\hbar} \>\xi_{2n} \> Y_{42n}\> X_{23n}+ $$
$$ + e^{- \imath E_{3n} t/\hbar} \> \xi_{3n}\> Y_{43n}\> X_{33n} + e^{- \imath E_{4n} t/\hbar}\> \xi_{4n} \> Y_{44n}\> X_{43n},
$$
$$
Z_{44,n} = e^{-\imath E_{1n} t/\hbar}\> \xi_{1n}\> Y_{41n}\> X_{14n} + e^{- \imath E_{2n} t/\hbar}\> \xi_{2n}\> Y_{42n}\> X_{24n}+ $$
$$+ e^{- \imath E_{3n} t/\hbar}\> \xi_{3n}\> Y_{43n}\> X_{34n} + e^{- \imath E_{4n} t /\hbar}\> \xi_{4n}\> Y_{44n}\> X_{44n}.$$

If the initial state of considered system is $|-,-,1\rangle$, the time dependent wave function  of the whole system takes the form
$$|\Psi(t)\rangle = Z_{14} |-,-,1\rangle + Z_{24} |+,-,0\rangle  + Z_{34} |-,+,0\rangle, \eqno{(10)}
$$
where
$$
Z_{13} =  \frac{1}{ \Omega}(\alpha^2 +  \cos\Omega t), \quad Z_{23} = -\frac{\imath}{\Omega} \sin\Omega t, \quad Z_{33} = - \frac{ \alpha}{ \Omega^2}(1-\cos\Omega t).
$$
At last, if the initial state  is $|-,-,0\rangle$, the time dependent wave function  will  evolve to
$$|\Psi(t)\rangle = e^{-\hbar \omega t} |-,-,0\rangle. \eqno{(11)}$$

Now we go back to the theme of this paper. If the initial state for
two atoms is (2), then using the equations (4)-(11) one can obtain the density operator for the whole system. Taking a partial trace over the heat bath variable one can obtain that the reduced atomic density operator in the two-atom basis evolves to
\begin{center} $\quad$
$$
\rho_A(t) = \left (
\begin{array}{cccc}\vspace{2mm}
\rho_{11} & \rho_{12} & \rho_{13} & \rho_{14}\\ \vspace{2mm}
\rho_{12}^* & \rho_{22} & \rho_{23} & \rho_{24}\\ \vspace{2mm}
\rho_{13}^* & \rho_{23}^* & \rho_{33} & \rho_{34}\\ \vspace{2mm}
\rho_{14}^* & \rho_{24}^* & \rho_{34}^* & \rho_{44}\\
\end{array} \right ).\eqno{(11)}$$
\end{center}
The matrix elements of (11) are given in Appendix.

For  two-qubit system described by the density operator $\rho_A(t)$, a measure of entanglement or negativity can be defined in
terms of the negative eigenvalues $\mu_i^-$ of partial transpose of the reduced atomic density matrix $\rho_{A}^{T_1}$ \cite{Peres},\cite{Horod}.
The  partial transpose of the reduced atomic density matrix (11) is
$$ \rho _{A} ^{T_{1} } (t)=\left(\begin{array}{cccc} {\rho_{11}} & {\rho_{12} } & {\rho_{13}^*} & {\rho_{23}^* } \\
{\rho_{12}^* } & {\rho_{22} } & {\rho_{14}^* } & {\rho_{24}^* } \\
 {\rho_{13}} & {\rho_{14} } & {\rho_{33} } & {\rho_{34} } \\
 {\rho_{23} } & {\rho_{24} } & {\rho_{34}^*} & {\rho_{44}} \end{array}\right).$$
The negativity is
$$\varepsilon= - 2 \sum\limits \mu_i^-.\eqno{(12)}$$
 When $\varepsilon= 0$  two qubits are
separable and $\varepsilon > 0$ means the atom-atom entanglement.
The case $\varepsilon= 1$ indicates maximum entanglement.

The results of calculations of entanglement parameter (12) for initial atomic states (2) are shown in Fig. 1-3. For all figures we put $\varphi_1=\varphi_2=0$.

\section{Results}

In Fig. 1 we plot the negativity as a function of $gt$ for coherent and three incoherent atomic states described by formula (2). The curves corresponds to   a fixed value of dipole strength  $\alpha = 0.5$ and   weak thermal two-mode field with $\bar n =0.01$. From these figure we can find that, firstly, entanglement can be induced by thermal field for all initial atomic state. The more  significant result is that the
atomic entanglement may be induced by thermal field when both the
atoms are prepared in their excited states. By contrast,   the
atomic entanglement does not  induced by thermal field when both atoms are trapped in cavity and prepared in  their excited states \cite{Kim}.
 Secondly, the maximum degree of  entanglement  is slightly enhanced owing to the atomic coherence comparing the curves in Fig.1(c)   and Fig.1(d).
With increasing of the mean photon number (see Figs.2(a) and Fg2(b)) the value of atom-atom negativity  decreases. But for  coherent states this decreasing is  much sharper.  Such entanglement behavior essentially differs from that  for model when both atoms are trapped in cavity and interact with a thermal field. In the last case  \cite{Hu2} the entanglement is greatly enhanced due to the initial
atomic coherence for thermal cavity with large mean photon numbers.
Fig. 3 displays how the evolution of entanglement  depends on the dipole-dipole strength   for $\bar n = 1$. From these figures, we find that entanglement revealed by the solid lines ($\alpha=0.1$) is
stronger than that by the dotted lines ($\alpha=1$) at most time. The value of atom–atom negativity is larger when the
dipole interaction between the atoms increases. This dependence is  much sharp for incoherent initial atomic state.

\begin{figure}[!h]
\begin{tabular}{c}
\mbox{(a)} \\
\includegraphics[scale=0.55]{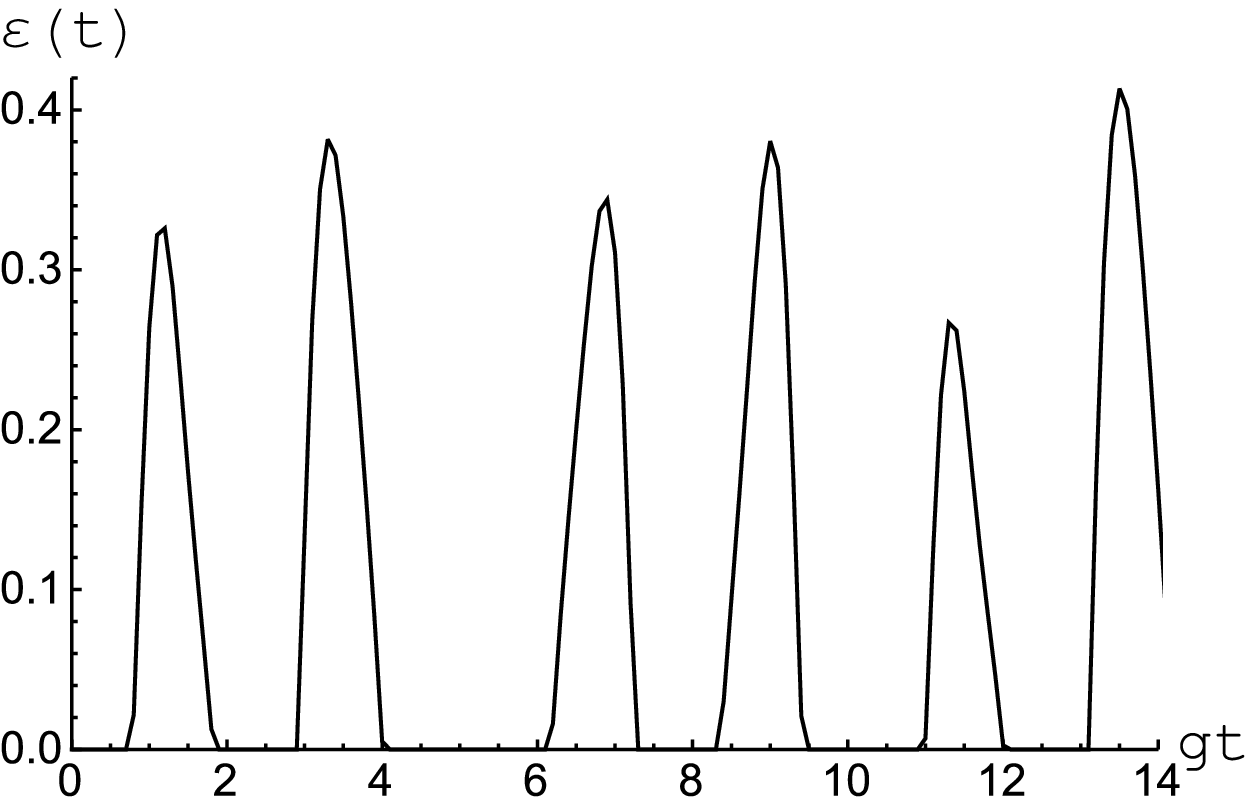}\\
 \mbox{(b)} \\
   \includegraphics[scale=0.55]{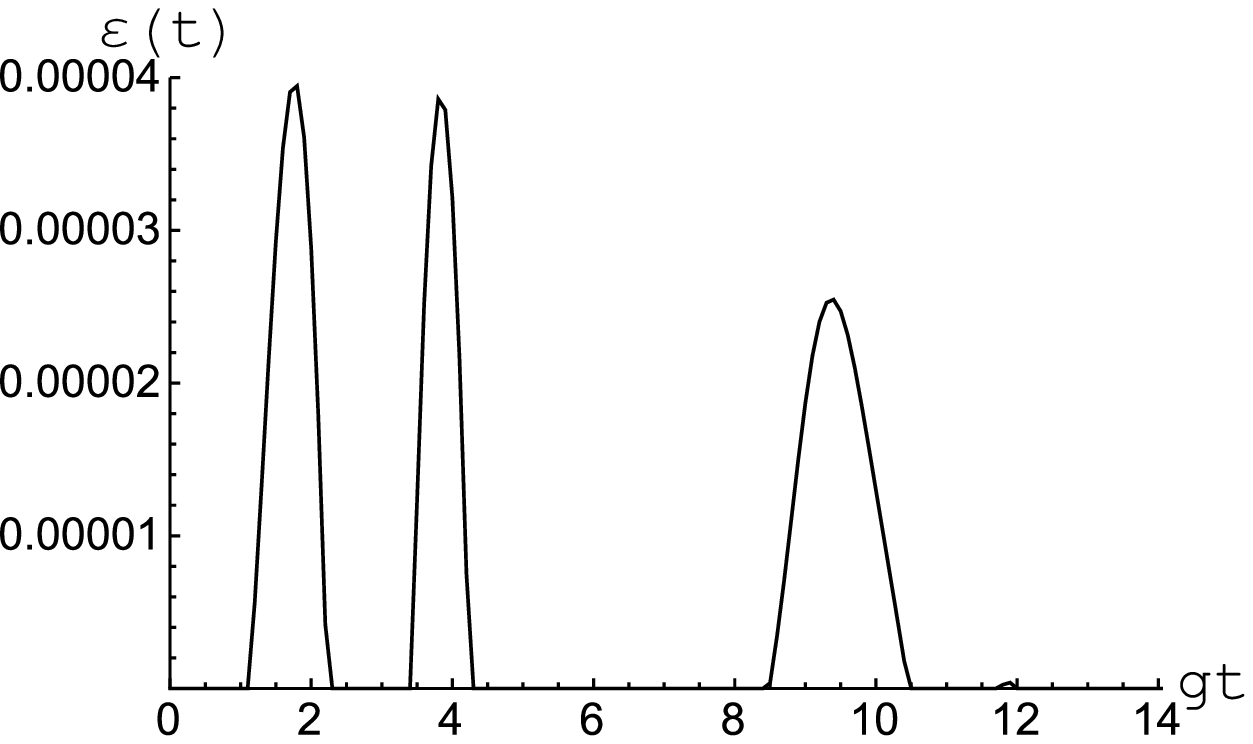}\\
\mbox{(c)} \\
\includegraphics[scale=0.55]{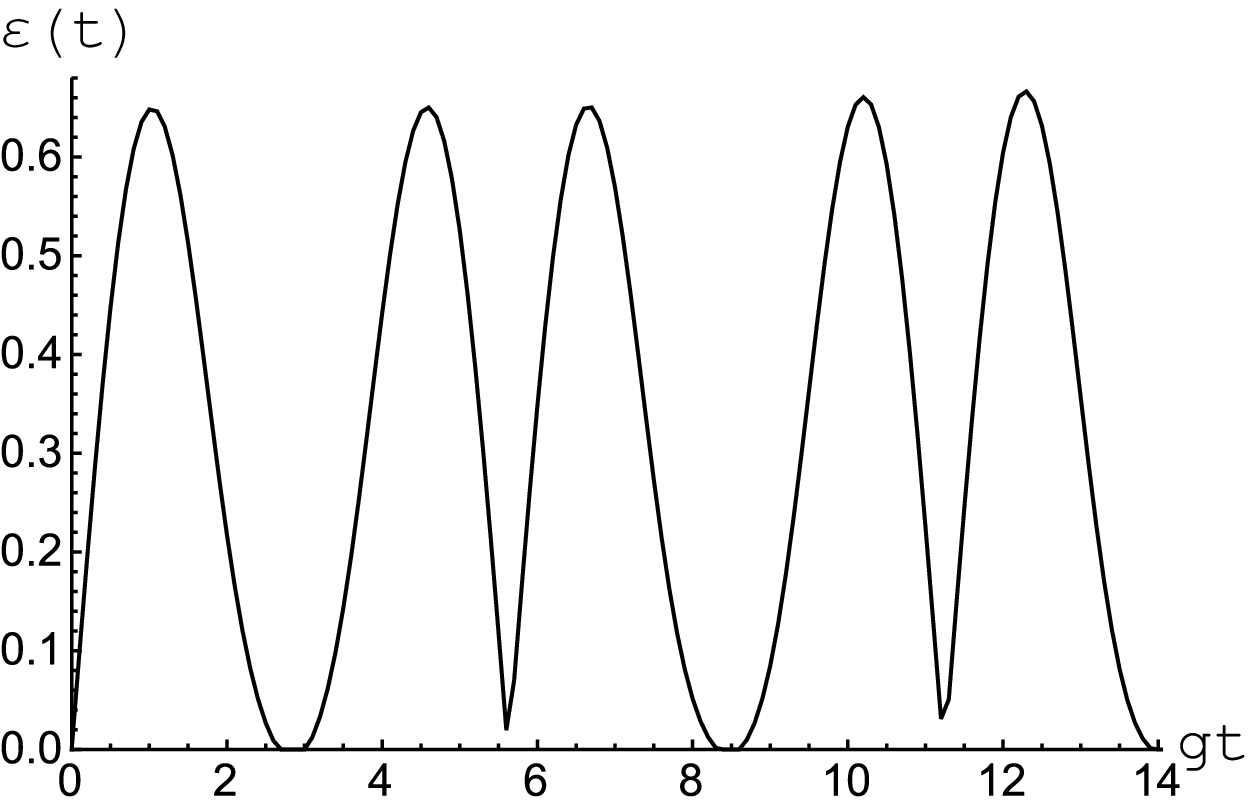}\\
  \mbox{(d)}
 \\  \includegraphics[scale=0.55]{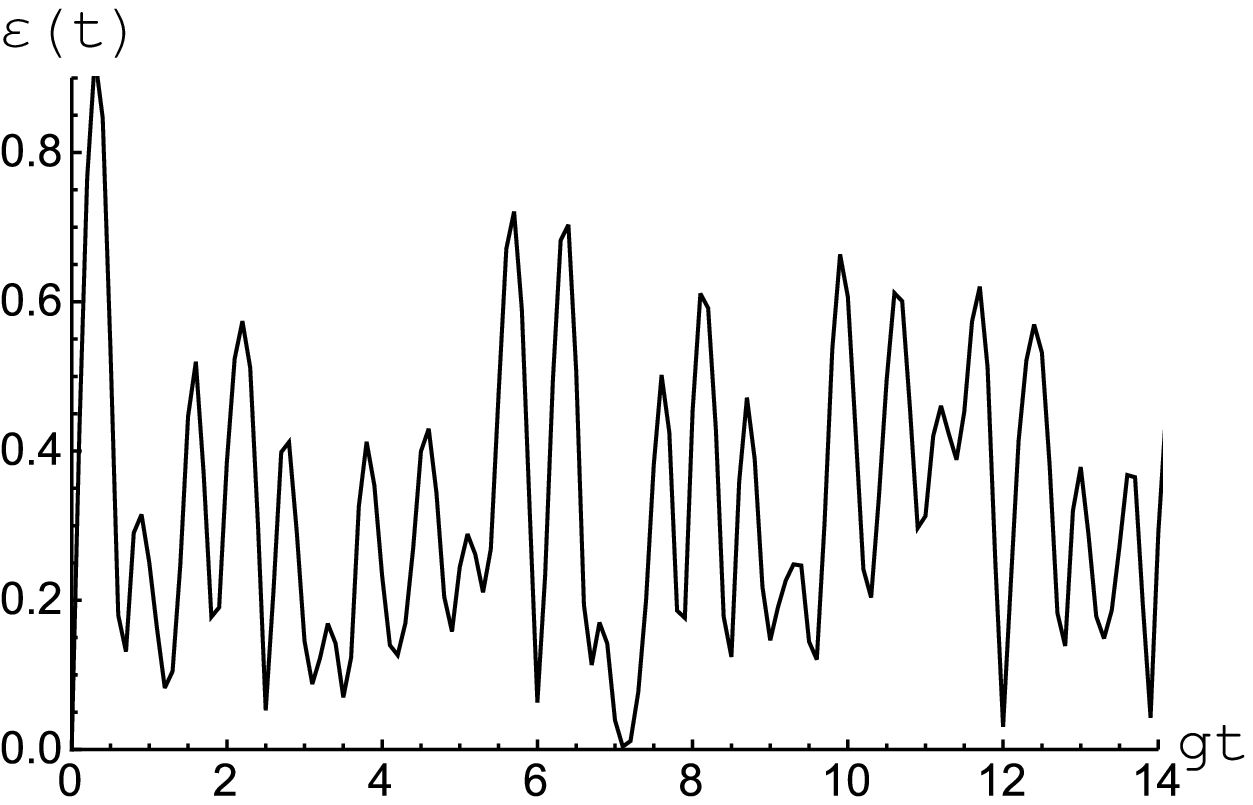} \\
\end{tabular}
\caption{The negativity as a function of $gt$  for the model  with  $\bar n = 0.01$ and  $\alpha = 0.5$. The initial atomic state:  a)   $|\Psi(0)\rangle_A= |+,-\rangle$ (a), $|\Psi(0)\rangle_A= |-,-\rangle$ (b), $|\Psi(0)\rangle_A= |+,+\rangle$ (c) and $|\Psi(0)\rangle_1=(1/\sqrt{2}(|+\rangle + |-\rangle), \quad   |\Psi(0)\rangle_2 = (1/\sqrt{2}(|+\rangle + |-\rangle) $ (d). }
\end{figure}

\begin{figure}[!h]
\begin{tabular}{c}
\mbox{(a)} \\
\includegraphics[scale=0.6]{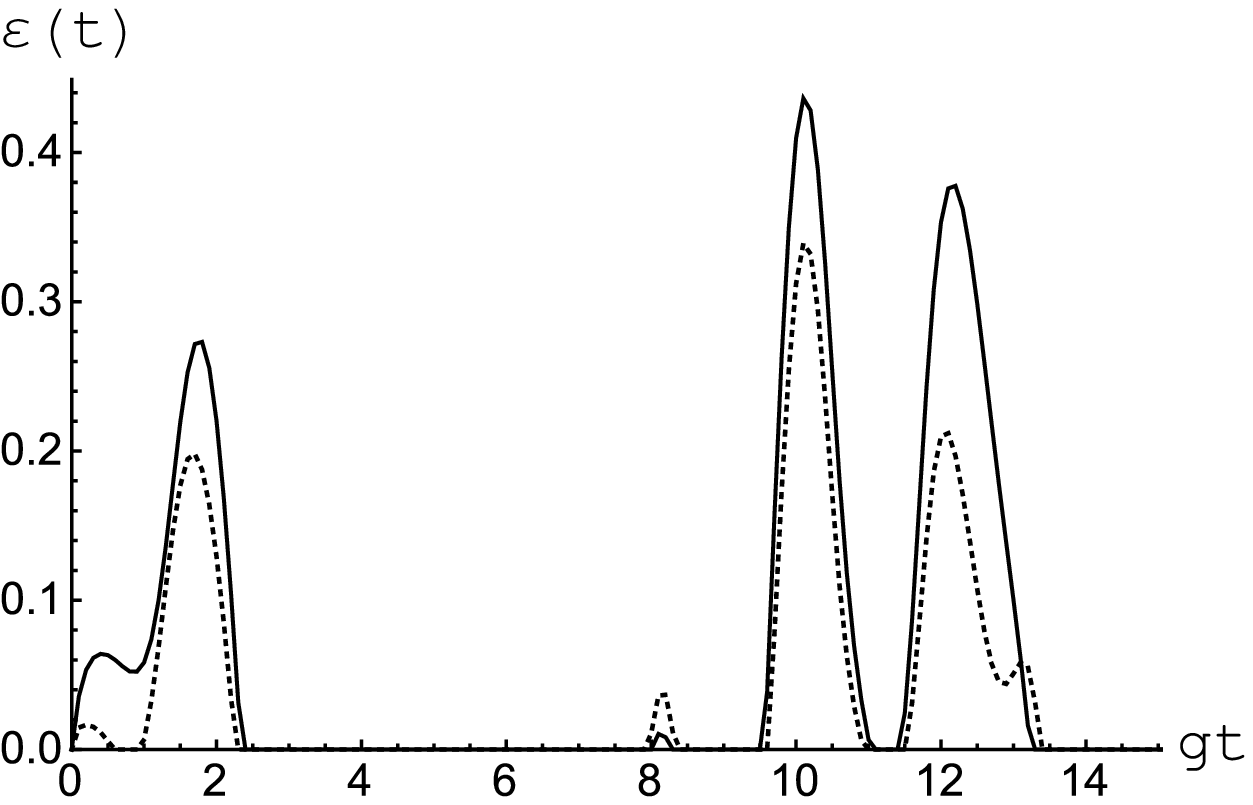}\\
 \mbox{(b)} \\
  \includegraphics[scale=0.6]{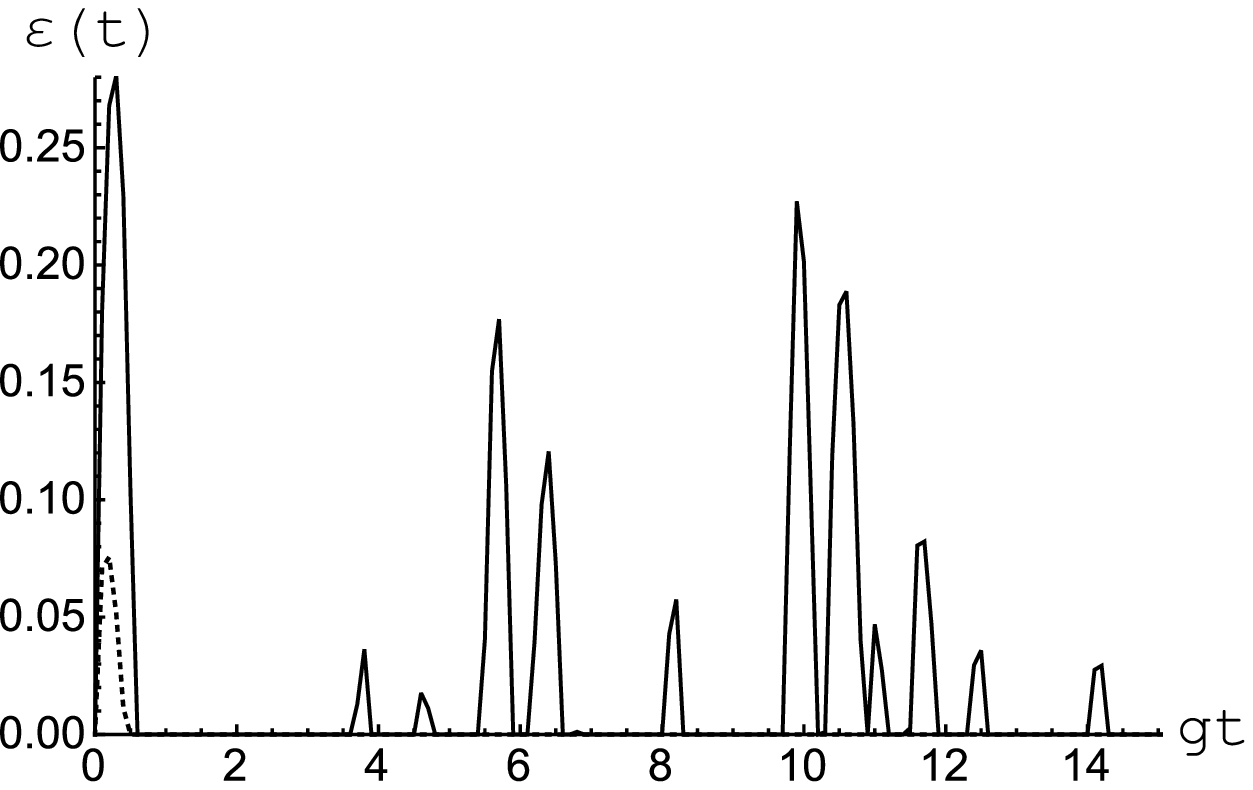}\\
\end{tabular}
\caption{The negativity as a function of $gt$  for the model  with  $\bar n = 1$ (solid) and $\bar n =2$ (dashed).
   The initial atomic state:  a)   $|\Psi(0)\rangle_A= |+,-\rangle$ (a) and $|\Psi(0)\rangle_1=(1/\sqrt{2}(|+\rangle + |-\rangle),\quad  |\Psi(0)\rangle_2 = (1/\sqrt{2}(|+\rangle + |-\rangle) $ (b). The dipole strength $\alpha = 0.5$ }
\end{figure}
\begin{figure}[!h]
\begin{tabular}{cc}
\mbox{(a)} \\
\includegraphics[scale=0.6]{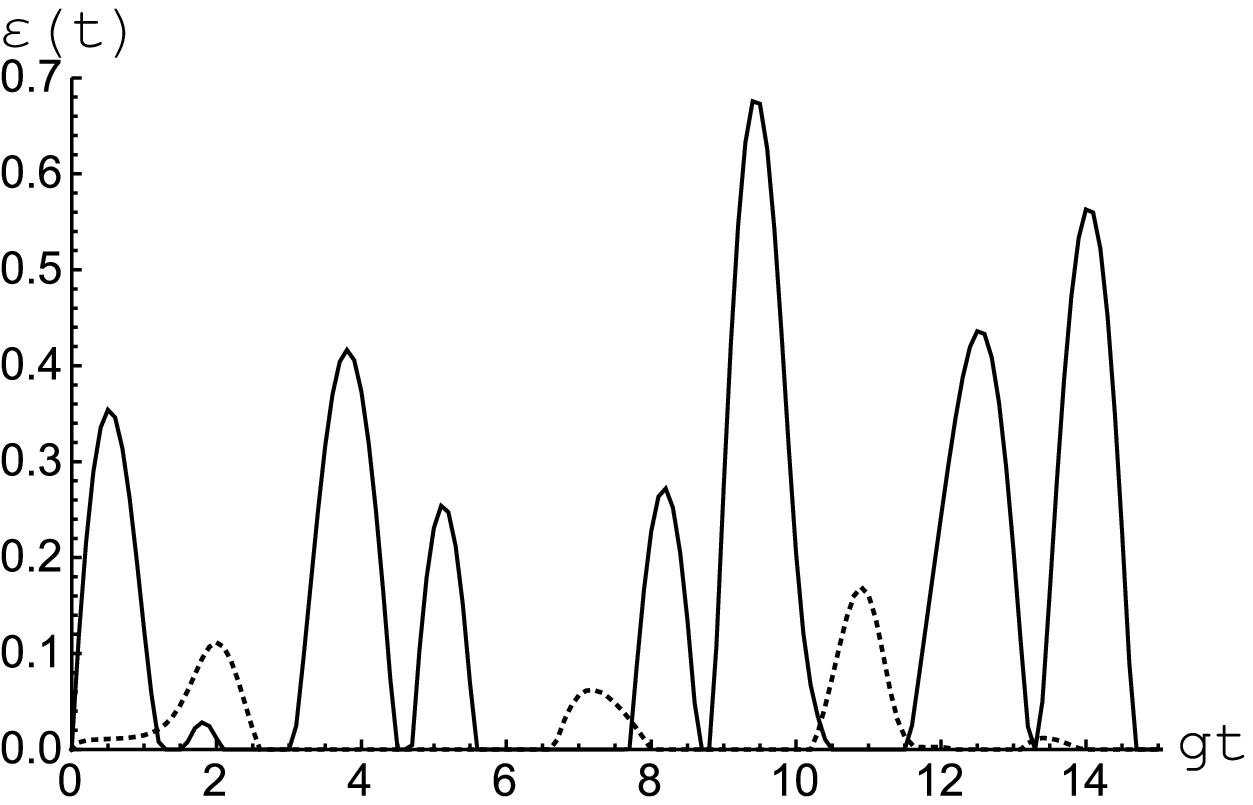} \\
 \mbox{(b)} \\   \includegraphics[scale=0.6]{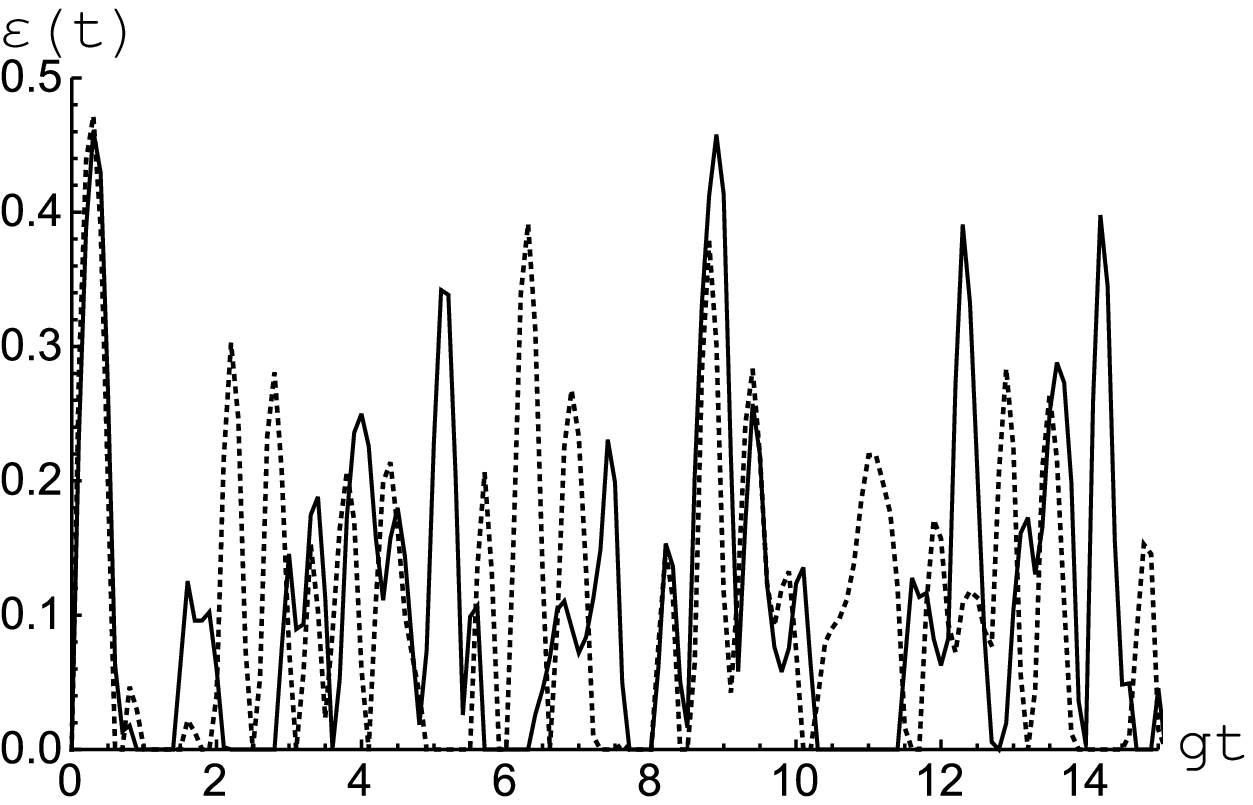}\\
\end{tabular}
\caption{The negativity as a function of $gt$  for the model  with  $\alpha = 0.1$ (solid) and $\alpha =1$ (dashed).
   The initial atomic state:  a)   $|\Psi(0)\rangle_A= |+,-\rangle$ (a) and $|\Psi(0)\rangle_1=(1/\sqrt{2}(|+\rangle + |-\rangle),\quad  |\Psi(0)\rangle_2 = (1/\sqrt{2}(|+\rangle + |-\rangle) $ (b). The mean photon number  $\bar n= 0.5$ }
\end{figure}

\section{Conclusion}

In this paper we have investigated the effect of the atomic coherence on the entanglement of two dipole coupled two-level  atoms when only one atom is trapped in a lossless cavity and  interacts with one-mode thermal field, and the other one can be spatially moved freely outside the cavity.    We have shown that thermal field can produce atom-atom entanglement
for all pure initial atomic states.  The results also  show that the atom-atom entanglement can be controlled by changing the system parameters, such as  the amplitudes of the polarized atoms, the mean photon numbers of  thermal field, and the  strength  of the dipole interaction (or distance between atoms).  We have derived that for considered model the dipole-dipole interaction can produce the appreciable amount of  entanglement only for small thermal noise if the atoms are  prepared in the coherent states.

\section*{Acknowledgments}

Work is carried out with financial support of the Ministry of Education and Science of the Russian Federation (the state assignment N 1394).

\section*{Appendix: Expressions for density matrix Elements in Eq. (11)}

The evident expressions for matrix elements in (11) are

$$\rho_{11}=|a|^2  \sum\limits_{n=0}  p_n |Z_{41,n}|^2 + |b|^2 \sum\limits_{n=1} p_n |Z_{42,n-1}|^2 +b c^*\sum\limits_{n=1} p_n Z_{42,n-1} Z_{43,n-1}^* +$$ $$
+  c b^* \sum\sum_{n=1} p_n  Z_{43,n-1} Z_{42,n-1}^* + |c|^2 \sum\limits_{n=1} p_n |Z_{43,n-1}|^2 + |d|^2  \sum\limits_{n=2} p_n  |Z_{44,n-2}|^2,$$
$$\rho_{12} = a b^*\sum\limits_{n=1} p_n Z_{41,n} Z_{22,n-1}^* +  a c^* \sum\limits_{n=1} p_n Z_{41,n} Z_{23,n-1}+ $$
$$+ b d^* \sum\limits_{n=2} p_n Z_{42,n-1} Z_{24,n-2} + c d^* \sum\limits_{n=2} p_n Z_{43,n-1} Z_{24,n-2}^*+ $$ $$ + p_1 (b d^* Z_{42,0} G_{24}^* +c d^* Z_{43,0} G_{24}^*)
+ p_0 (a b^* Z_{41,0} G_{22}^* +a c^* Z_{41,0} G_{23}^* ),$$
$$\rho_{13} = a b^*\sum\limits_{n=1} p_n Z_{41,n} Z_{32,n-1}+  a c^* \sum\limits_{n=1} p_n Z_{41,n} Z_{33,n-1}^* + $$
$$ +b d^*\sum\limits_{n=2} p_n Z_{42,n-1} Z_{34,n-2}^* +c d^*\sum\limits_{n=2} p_n Z_{43,n-1} Z_{34,n-2}^*+$$
$$ +p_1( b d^* Z_{42,0} G_{34}^* + c d^* Z_{43,0} G_{34}^*) + p_0 (a b^* Z_{41,0} G_{32}^* +a c^* Z_{41,0} G_{33}^*),$$
$$\rho_{14} = a d^*\sum\limits_{n=2} p_n Z_{41,n} Z_{14,n-2}^*+ p_1 a d^* Z_{41,1} G_{14}^* + p_0 a d^* Z_{41,0},$$
$$\rho_{22}=|a|^2 \sum\limits_{n=0} p_n |Z_{21,n}|^2 +|d|^2 \sum\limits_{n=2} p_n |Z_{24,n-2}|^2 + $$
$$ +b c^* \sum\limits_{n=1} p_n Z_{22,n-1} Z_{23,n-1}^* +  c b^* \sum\limits_{n=1} p_n Z_{23,n-1} Z_{22,n-1}^*+$$
$$ + |c|^2 \sum\limits_{n=1} p_n |Z_{23,n-1}|^2 +|b|^2 \sum\limits_{n=1} p_n  |Z_{22,n-1}|^2 + $$
$$ +p_1 |d|^2 |G_{24}|^2 +p_0 (|b|^2 |G_{22}|^2 + b c^* G_{22} G_{23}^*) + p_0 (c b^* G_{23} G_{22}^*  |c|^2 |G_{23}|^2),$$
$$\rho_{23} = |a|^2  \sum\limits_{n=0} p_n Z_{21,n} Z_{31,n}^* +|d|^2 \sum\limits_{n=2} p_n Z_{24,n-2} Z_{34,n-2}^*+$$
$$ +b c^* \sum\limits_{n=1} p_n Z_{22,n-1} Z_{33,n-1}^*+ c b^* \sum\limits_{n=1} p_n Z_{23,n-1} Z_{32,n-1} +$$
$$ + |c|^2 \sum\limits_{n=1} p_n Z_{23,n-1} Z_{33,n-1}^* + |b|^2 \sum\limits_{n=1} p_n Z_{22,n-1} Z_{32,n-1}^*+ $$
$$ +p[1] |d|^2 G_{24} G_{23}^* + p[0](|b|^2 G_{22} G_{32}^* + b c^* G_{22} G_{33}^* + c b^*
G_{23} G_{32}^* +|c|^2 G_{23} G_{33}^*), $$
$$\rho_{33}=|a|^2  \sum\limits_{n=0} p_n |Z_{31,n}|^2 + |b|^2 \sum\limits_{n=1} p_n |Z_{32,n-1}|^2 +b c^* \sum\limits_{n=1} p_n Z_{32,n-1} Z_{33,n-1}^* +$$
$$ + c b^* \sum\limits_{n=1} p_n Z_{33,n-1} Z_{32,n-1}^* + |c|^2 \sum\limits_{n=1} p_n  |Z_{33,n-1}|^22 + |d|^2 \sum\limits_{n=2} p_n |Z_{34,n-2}|^2+ $$
$$+p_1 |d|^2 |G_{34}|^2 + p_0 (|b|^2 |G_{32}|^2 + b c^* G_{32} G_{33}^* + c b^* G_{33} G_{32}^2 +|c|^2 |G_{33}|^2,$$
$$\rho_{34} = a b^* \sum\limits_{n=1} p_n Z_{31,n} Z_{12,n-1} +  a c^*\sum\limits_{n=1} p_n Z_{31,n} Z_{13,n-1} + $$
$$ +b d^* \sum\limits_{n=2} p_n Z_{32,n-1} Z_{14,n-2}^* + c d^* \sum\limits_{n=2} p_n Z_{33,n-1} Z_{14,n-2}^*+$$
$$+ p_1 (b d^* Z_{32,0} G_{14}^* +c d^* Z_{33,0} G_{14}^*) +  p_0(a b^* Z_{31,0} G_{12}^* + a c^* Z_{31,0} G_{13}^*)+$$
$$ +p_0(b d^* G_{32} + c d^* G_{33},$$
$$\rho_{44}=|a|^2  \sum\limits_{n=0} p_n |Z_{11}|^2 + |b|^2  \sum\limits_{n=1} p_n |Z_{12,n-1}|^2 + b c^* \sum\limits_{n=1} p_n Z_{12,n-1} Z_{13,n-1}^* + $$
$$ +  c b^* \sum\limits_{n=1} p_n Z_{13,n-1} Z_{12,n-1}^* + |c|^2 \sum\limits_{n=1} p_n |Z_{13,n-1}|^2 +|d|^2 \sum\limits_{n=2} p_n |Z_{14,n-2}|^2 + $$
$$ + p_1 |d|^2  |G_{14}|^2 + p_0(|b|^2 |G_{12}|^2 + b c^* G_{12} G_{13}^*+ c b^* G_{13} G_{12}^* +|c|^2 |G_{13}|^2 + |d|^2),
$$
where $$ a =\cos\theta_1 \cos\theta_2,\quad b = \cos\theta_1 \sin\theta_2 e^{\imath \varphi_2},\quad c = \cos\theta_2 \sin\theta_1 e^{\imath \varphi_1},\quad
d = \sin\theta_1 \sin\theta_2 e^{\imath (\varphi_1+\varphi_2)}.$$

\end{document}